\RequirePackage{lineno}
\documentclass[aps,prb,preprint,superscriptaddress]{revtex4}

\usepackage{graphicx}
\usepackage{bm}
\usepackage{amsmath}
\usepackage{amssymb}
\usepackage{hyperref}
\usepackage[version=3]{mhchem}
\usepackage{float}

\begin{document}

\title {Mechanisms governing phonon scattering by topological defects in graphene nanoribbons}

\author{Ziming Zhu, Xiaolong Yang}
\address{Frontier Institute of Science and Technology, Xi'an Jiaotong University, Xi'an 710049, P. R. China}
\author{Mingyuan Huang}
\address{Department of Physics, South University of Science and Technology of China, Shenzhen, P. R. China}
\author{Qingfeng He}
\address{Department of Civil Engineering, Hunan University, Changsha 410082, P. R. China}
\author{Guang Yang}
\address{Electronic Materials Research Laboratory, Key Laboratory of the Ministry of Education and International Center for Dielectric Research, Xi'an Jiaotong University, Xi'an 710049, P. R. China}
\author{Z. Wang}
\address{Frontier Institute of Science and Technology, State Key Laboratory for Mechanical Behavior of Materials, Xi'an Jiaotong University, 710054, Xi'an, P. R. China.}
\email{wzzhao@yahoo.fr}


\begin{abstract}
Understanding the phonon scattering by topological defects in graphene is of particular interest for the thermal management in graphene-based devices. We present a study at quantifying the roles of different mechanisms governing defect phonon scattering by comparing the effects of ten different defect structures using molecular dynamics. Our results show that the phonon scattering is mainly influenced by mass density difference with general trends governed by the defect formation energy, with typical softening behaviors in the phonon density of state. The phonon scattering cross-section is found to be far larger than that geometrically occupied by the defects. We also show that the lattice thermal conductivity can be reduced by a factor up to $30$ in presence of grain boundaries formed by these defects.
\end{abstract}

\maketitle
\section{Introduction}

Accurate heat flow manipulation is of fundamental challenge for many graphene-based devices including electronic devices and nanoelectromechanical systems (NEMS).\cite{Pao2012,Li2012} Recent experiments showed the existence of topological defects\cite{Hashimoto2004,Meyer2008a,Pop2012} exhibit significant influence on thermal transport properties of graphene.\cite{Huang2013,Morooka2008,Mortazavi2013,Wang2012,Xie2011,Yeo2012,Zhang2011,Jiang2011,Zhao2015}, and thus offer possibility to control graphene thermal conductivity by nanostructuring.\cite{Lindsay2010} For instance, Haskins \textit{et al.} proposed control of graphene thermal and electronic transport by introducing vacancy and Stone-Wales (SW) defects in graphene nanoribbons (GNRs).\cite{Haskins2011} Ng \textit{et al.} reported that the presence of the SW defects can decrease graphene thermal conductivity by more than 50\%.\cite{ng2012} Results of Tan and co-workers showed that the SW defect effect on thermal conduction in armchair-oriented graphene GNRs is stronger than that in zigzag-oriented GNRs.\cite{tan2013} Hao \textit{et al.} further studied the dependence of thermal conductivity upon the SW defect concentration.\cite{Hao2011}

Although the above results are revealing, most of recent studies were restored to one or two types of typical lattice defects such as the SW ones, a comprehensive comparison between different types of defect is absent, thus, the understanding on fundamental mechanisms governing phonon scattering by graphene topological defects remains fragmented. On the other hand, the structures of graphene lattice defects may depart considerably from one to others.\cite{Banhart2011} For instance, experimentalists have shown evidence for C-8 rings in graphene.\cite{Lahiri2010} Also, the existence of thermally stable point vacancies in graphene has been reported by a high-resolution transmission electron microscopy study.\cite{Hashimoto2004} The existing conditions for these defects are not clear up to date despite of their potential importance for graphene electronic and thermal transport properties.

\section{Method}
With the above motivations, it is crucial to understand the condition for existing different types of lattice defects and the correlation between the defect physical properties and graphene thermal conductivity. Here we search for possible atomistic configurations of lattice defects synthesized from different annealing processes, and calculate their influence on the graphene lattice thermal conductivity using molecular dynamics. In our simulations, defects are inserted into GNRs by either removing atoms or rotating neighboring C-C bonds in the honeycomb lattice, as shown in Fig.\ref{fig:1}. Molecular dynamics (MD) are then performed to find the stable atomistic configurations of the defective GNRs.\cite{wang2011prb,wangprl2009,Yang2015} We use free boundary condition in three orthogonal directions. The system is relaxed at different temperatures for $0.5$ ns with zero pressure using Nose-Hoover thermostat, before a stabilized configuration can be recorded. The inter-atomic potential used in our MD is the adaptive intermolecular reactive empirical bond-order hydrocarbon (AIREBO) model, in which the total inter-atomic energy is a collection of that of individual bonds.\cite{Stuart2000} Many-body effects are explicitly included by introducing a bond-order function, in which the influences of the bond angle, bond conjugation and atomic dihedral angle are taken into account. This approach showed accuracy in describing interactions between $sp^{2}$ carbon atoms in our previous studies.\cite{wang2011prb,wangprl2009}

The lattice thermal conductivity of defective GNRs is calculated by non-equilibrium molecular dynamics (NEMD), in which a temperature difference is imposed between the two fixed edges of a suspended GNR. The temperature gradient $\Delta T$, as well as the heat current $J$ are recorded in analogous to the typical experimental setup for measuring the thermal conductivity, which is defined by the Fourier's law, $\kappa ={J l}/{\Delta T wt}$, where $l$ and $w$ stand respectively for the length and width of thermal conduction part, and a value of $0.335$nm is assigned to $t$ as the inter-planar spacing of graphite stacking. Our NEMD simulations consist of three steps: A Berendsen thermostat is used to help the system reaching equilibrium at room temperature in the first simulation phase; The temperatures in the heat source and sink are then controlled to reach the settled values ($320$ and $280$ K respectively); The GNR center progressively reaches a steady state in $10^{6}$ simulation steps. The measured quantities are averaged over the third step of further $10^{6}$ steps. It is noted that we do not fix the ribbon in the direction normal to the graphene plane for getting closer to a real graphene, the ribbon geometry is thus not perfectly flat but with distortion to the third dimension. One to five randomly placed defects per GNR are considered in the calculation with zigzag edge geometry along the direction of heat flow.

\section{Results and Discussions}

\begin{figure}[htp]
\centerline{\includegraphics[width=10cm]{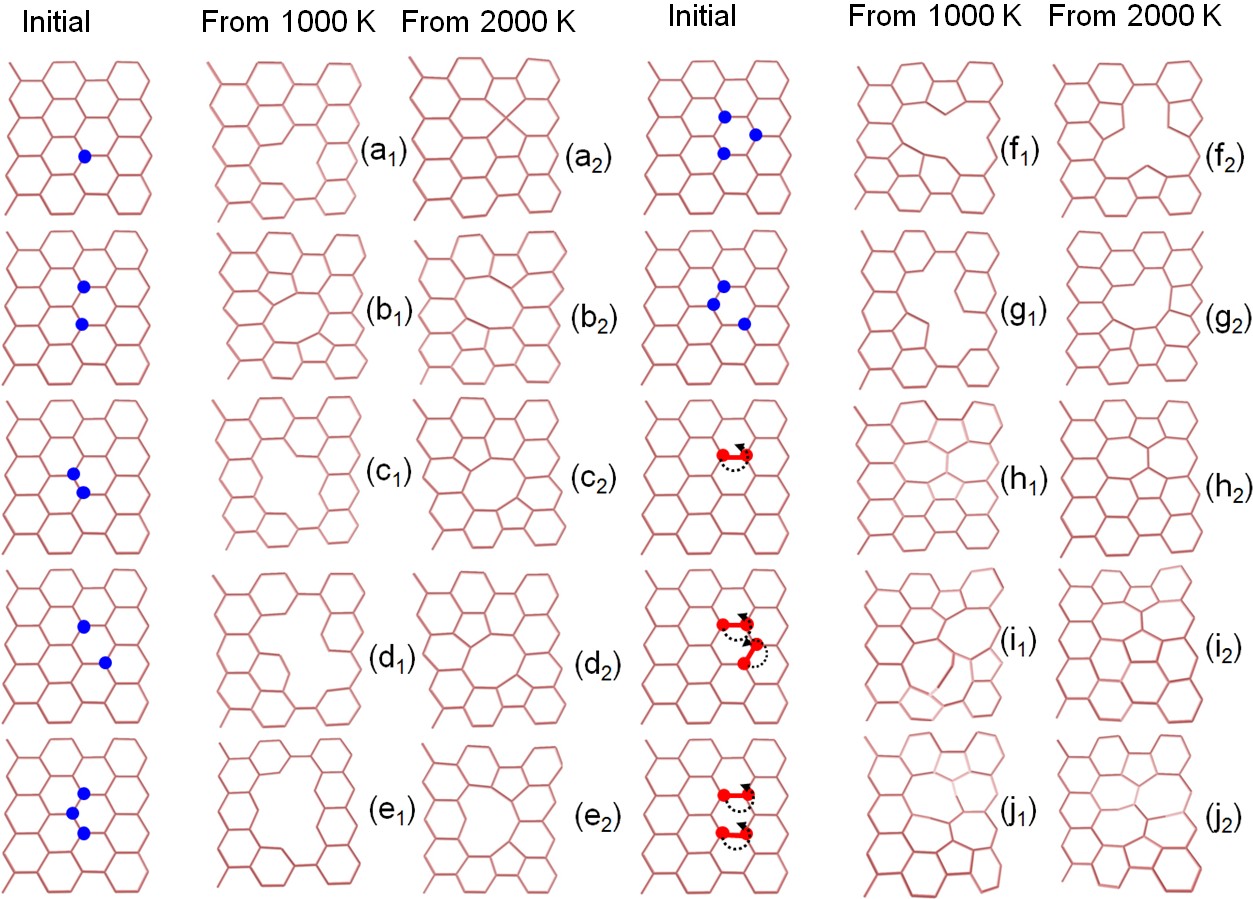}}
\caption{\label{fig:1}
Snapshots of lattice defects formed by point vacancies (a)-(g) or bond rotations (h)-(j) at room temperature. Atomistic configurations are obtained by annealing from $1000$K (subscript 1) or $2000$K (subscript 2).}
\end{figure}

Since graphene grows through very high temperature in its typical synthesis process, we simulate several different structures of defective graphene at $1000$K and $2000$K, before cooling down to the room temperature. The ribbon width is set to be large enough ($20$nm) in order to reduce the edge effects. Our results show that most point vacancies are relatively stable at about $1000$K [Fig.\ref{fig:1} (a), (c), (d), (e), (f) and (g)], while reform to other more stable structures at above $2000$K. This is consistent with the high-resolution transmission electronic microscopy observation reported by Hashimoto and co-worker.\cite{Hashimoto2004} Figs.\ref{fig:1} (h-j) also shows that the well-known Stone-Wales defects are indeed the most energetically favorable structures formed by the bond rotation, in agreement with that previously suggested by Meyer and coworkers.\cite{Meyer2008a} Furthermore, the a 5-8-5 carbon rings defect structures formed through the reconstruction of di-vacancy [Fig.\ref{fig:1} (b), (c) and(d)] is also consistent with those previously reported.\cite{Botello-Mendez2011} We notice that this structure exhibits significant distortion to the direction normal to the graphene plane. Since hybridization of the $\pi$ orbital is supposed to be markedly modified by such large lattice distortion, important change in the electronic properties can be expected.\cite{Morpurgo2006} This effect will be particularly important in multi-layered graphene, due to modified van der Waals interactions and the possibility of spontaneously-formed covalent bonds with neighboring layers by thermal fluctuation.\cite{Nilsson2006}

\begin{figure}[htp]
\centerline{\includegraphics[width=11cm]{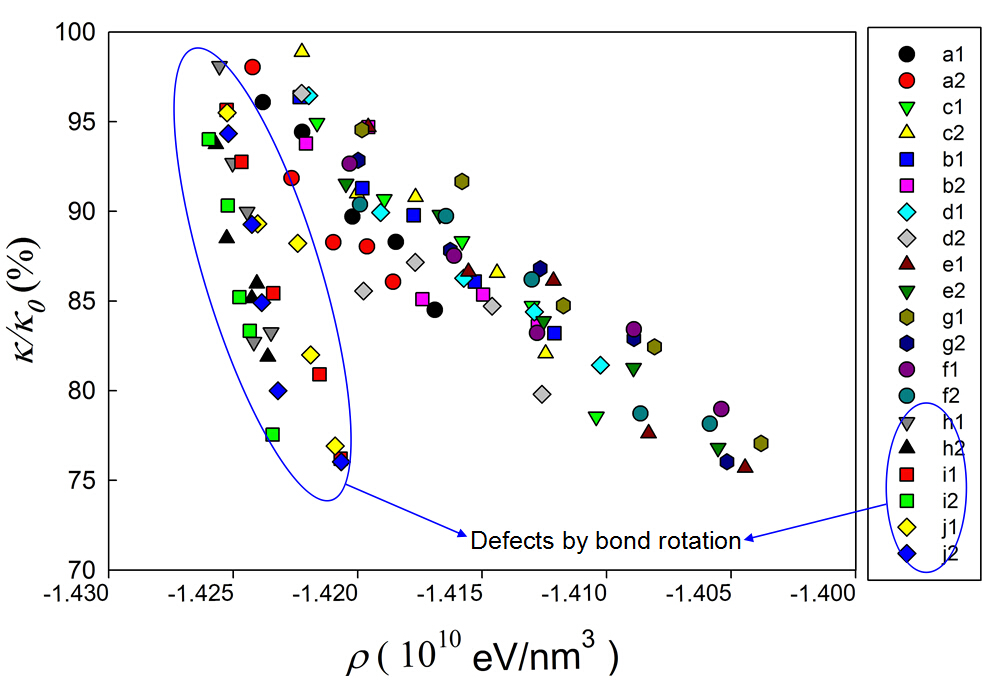}}
\caption{\label{fig:3}
Graphene lattice thermal conductivity as a function of the density of formation energy. The thermal conductivity $\kappa$ of defective GNRs is normalized with respect to that of a pristine GNR of the same size $\kappa_{0}$.}
\end{figure}

Physically speaking, there should be three major dependence of the phonon scattering by lattice defects: 1. The local bonding energy/force change; 2. The mass difference caused by replacing or removing atoms; 3. The defective cross-section area or width normal to the heat flow. We first plot the lattice thermal conductivity as a function of the atomic bonding energy density ($\rho$) in Fig.\ref{fig:3}, in which an inverse proportionality can be clearly seen. We find that the group of defects formed by bond rotation exhibits a clearly different dependency with the defects formed by point vacancies. This suggest that the mass density difference plays an important role besides the local energy change, while the effect of the cross section area is not as significant as the above-discussed mechanisms. No significant difference is found between the defect structures annealed from two different temperatures. The e1, g1, j1 (or e2, g2, j2) defects are found to reduce thermal conductivity the most, far beyond the limit of the typical SW defects.

\begin{figure}[htp]
\centerline{\includegraphics[width=11cm]{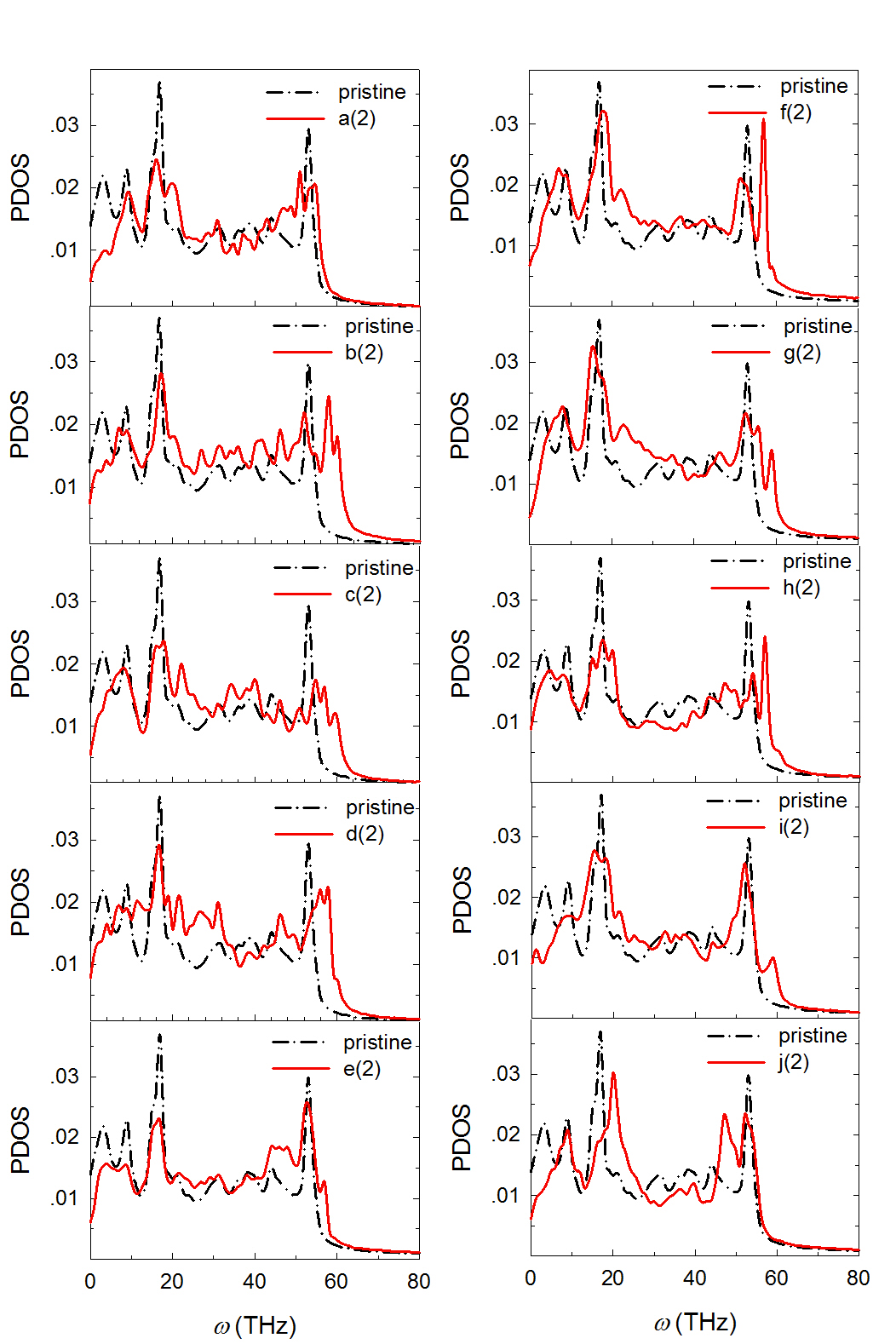}}
\caption{\label{fig:4a}
Phonon density of state for the lattice defects shown in Fig.\ref{fig:1} annealed from $2000$K.}
\end{figure}

To further understand the defect effect on phonon transport, we compute the phonon density of state (PDOS) for each type of the above-discussed defects that are shown in Fig.\ref{fig:4a}. It can be seen that, comparing with pristine graphene, the phonon spectrums of defective graphene structures have a trend to shift toward higher frequency. The phonon mode around 18 THz and 55 THz are broadened when the lattice defects are introduced. These indicate a typical phonon softening behavior. In general, the phonon lifetime $\tau$ can be expressed as $\tau = 0.5 {L}^{-1}$, where $L$ is the half-width at half-maximum.\cite{Mu2014,Wang2011a,wang2010f} Therefore, broadening of phonon modes indicates reduction in the phonon lifetime and thus leads to the thermal conductivity decrease. This is in good agreement with Ref.\cite{Wang2014,Wei2011}. This PDOS peak broadening effect may also be explained as the phonon scattering enhancement by the mass density difference.\cite{lin2014a}

\begin{figure}[htp]
\centerline{\includegraphics[width=11cm]{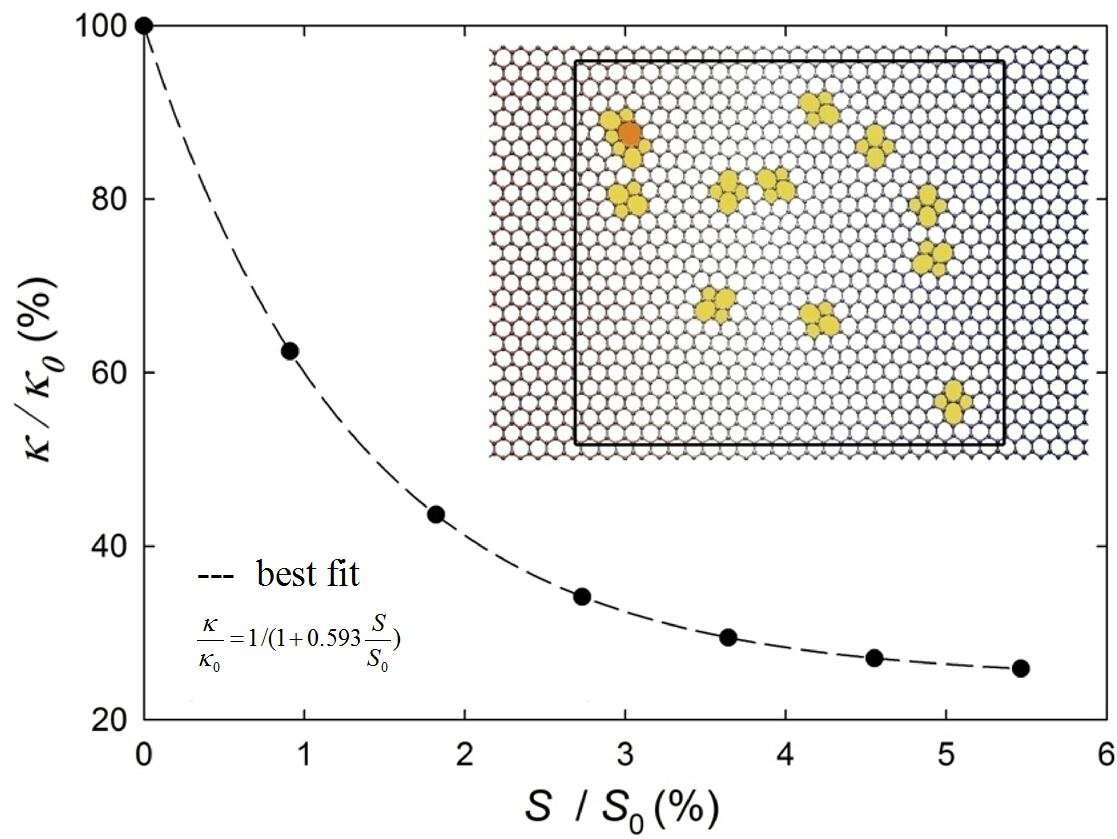}}
\caption{\label{fig:4}
$\kappa/\kappa_{0}$ as a function of the defected surface occupation $S/S_{0}$. The dashed line shows a best-fit curve. Inset: Snapshot of a simulated sample.}
\end{figure}

To show a quantitative correlation between the defect concentration and the thermal conductivity, in Fig.\ref{fig:4} we plot the thermal conductivity ratio of the different defects as a function of the defect concentration. We see that $\kappa/\kappa_{0}$ exhibits an exponential-like decay at increasing defect number, this is in agreement with the results reported by Hao and co-workers.\cite{Hao2011} Similar thermal conductivity variation trends with respect to defects concentration were also reported previously.\cite{Sevik2011,Zhang2012, Khosravian2013} It should be mentioned that our simulations may not provide the accurate absolute value of the graphene thermal conductivity since the sample size is far smaller than the phonon mean free path in graphene.\cite{Balandin2011}

\begin{figure}[htp]
\centerline{\includegraphics[width=11cm]{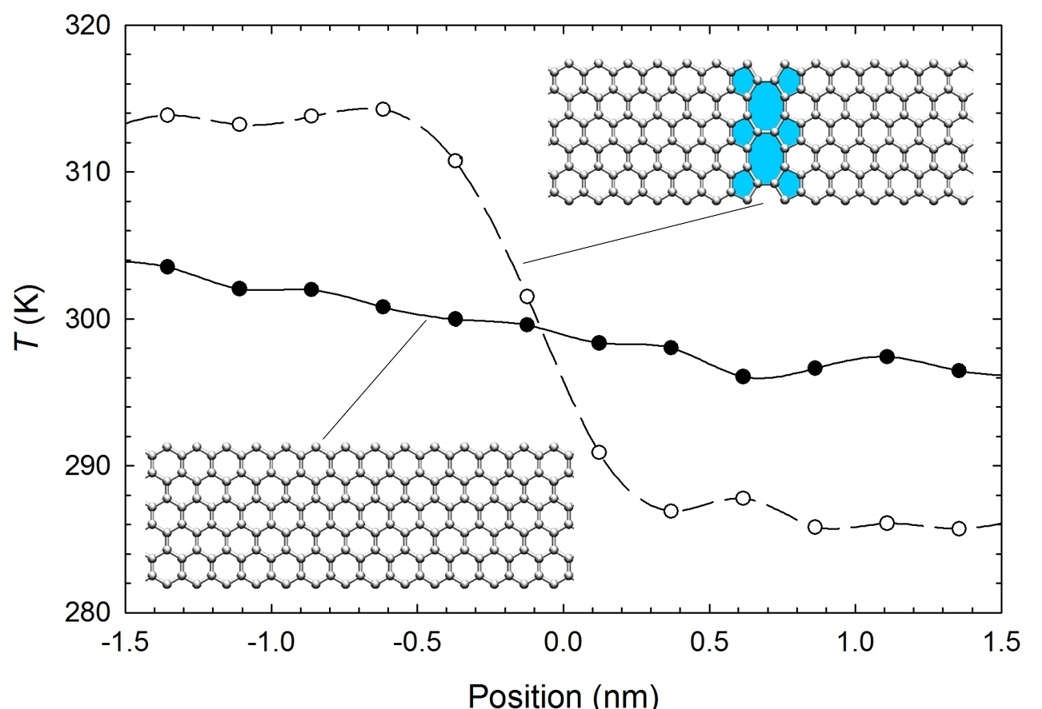}}
\caption{\label{fig:5}
Temperature profile along the longitudinal direction of both a pristine GNR and one with the cross section completely blocked by 5-8-8-5-ring defects (ribbon width $4.84$nm). Inset: Pristine (\textit{bottom left}) and defective (\textit{up right}) structures.}
\end{figure}

The lattice defects can connect with each other to form line defects, which are widely distributed as grain boundaries in graphene synthesized through chemical vapor deposition. Giving their importance we extended our NEMD simulations on a narrow GNR with aligned 5-8-5 defects placed at the ribbon center. Our results in Fig.\ref{fig:5} shows a sharp temperature decrease crossing the line defects, resulting in a extremely high thermal resistance.\cite{Lee2013} The thermal conductivity is found to be reduced by a factor of about $29.7$. We note that, theoretical speaking, the sharp jump in the temperature profile should be weakened if the ribbon gets smaller, due to the corresponding decrease of phonon mean free path.\cite{Wang2011a} However, this effect should not be significant in our case since the phonon mean free path is supposed to be much larger than the ribbon's longitudinal dimension.

\begin{figure}[htp]
\centerline{\includegraphics[width=11cm]{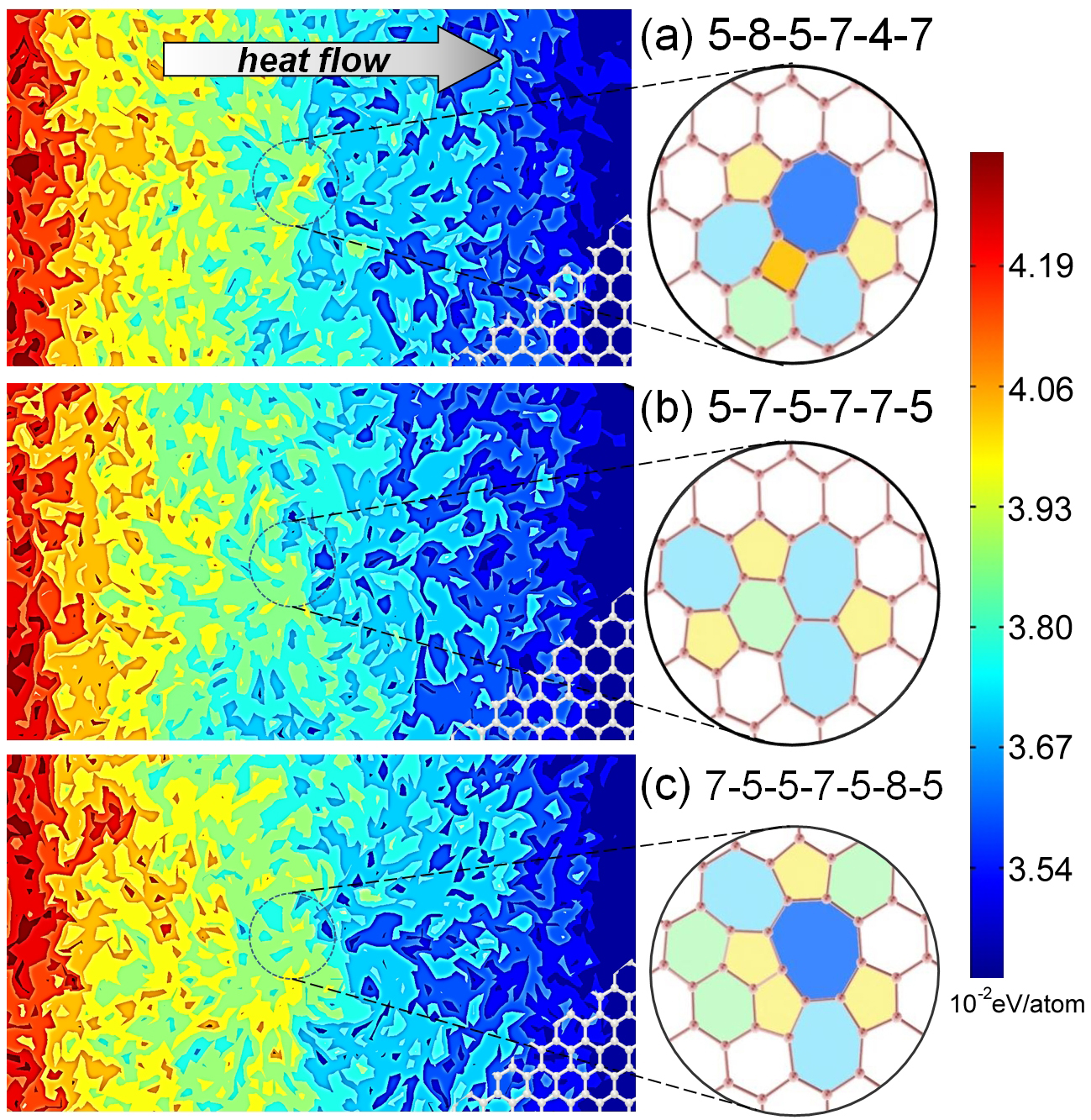}}
\caption{\label{fig:6}
Thermal energy distributions in three GNRs containing different types of defects. \textit{Left panels:} The color scale corresponds to the dynamic energy density. \textit{Right panels:} Different polygon lattices shown in different colors.}
\end{figure}

To illustrate the similarity and difference between heat flow in graphene with defect phonon scattering and fluid flow passing an obstacle, in Fig.\ref{fig:6} we depict the thermal energy distribution in GNRs with three different lattice defects. We see that the thermal energy distribution at most parts of the GNR surface is modified with a surface area much larger than that geometrically occupied by the lattice defect. This is because phonons reflected by the defect collides further with phonons propagating in other directions, and finally form a large scattering pattern. In comparison, we see that the C-5-7-5-7-7-5 rings scatter the less with phonons [Fig.\ref{fig:6} (b)], and give the most homogeneous heat distribution. With the highest-energy and distortion to the third dimension, the C-5-8-5-7-4-7 rings exhibit the strongest phonon scattering, and generates a high-temperature spot [Fig.\ref{fig:6} (a)]. The phonon scattering surface by this defect is particularly large with thermal energy distribution markedly modified at more than half of the GNR surface. The C-7-5-5-7-5-8-5 rings exhibit the largest distorted surface area in Fig.\ref{fig:6} (c), creating the broadest phonon scattering region near the heat source. We note that the distance between the edges and the defects could play an important role to change the mean free path of phonon, which would further impact on the thermal conductivity.\cite{Hu2009}

\section{Summary}

In summary, we simulate phonon scattering in GNRs by ten different defect structures. Our results show that the local energy gradient is the key to reduce the thermal conductivity, while the mass density difference plays a leading role. Phonon density of state analyses show a typical phonon softening behavior induced by the topological defects. We also show that the lattice thermal conductivity can be reduced by a factor up to $30$ in presence of grain boundaries. The scattering pattern is found to be far larger than that geometrically occupied by the defects. The existence of these lattice defects including C-5, C-7, and C-8 rings can be expected to likewise explain experimentally measured thermal conductivities lower than the ballistic limit \cite{Balandin2011,Shakouri2006}, and could probably suggest new lines of graphene microscopy research for experiments.


\section*{References}


\end{document}